\documentclass[a4paper]{article}

\newcommand{\sect}[1]{\section{#1}\setcounter{equation}{0}}
\newcommand{\bidx}[2]{\underset{#2}{#1}}

\usepackage{amsmath}
\usepackage{amsfonts}

\begin{document}

\title{On irreducible partials of Ricci tensor traceless part
in finite space-time region in GR}
\author{Yu. Semenov\footnote{Odessa National Polytechnical University,
Odessa, Ukraine, e-mail: yury@paco.net}}
\date{October 28, 2002}
\maketitle

\begin{abstract}

Riemann tensor irreducible part $E_{iklm} = \frac{1}{2} ( g_{il}S_{km} + g_{km}S_{il}
- g_{im}S_{kl} - g_{kl}S_{im} )$ constructed from metric tensor $g_{ik}$ and
traceless part of Ricci tensor $S_{ik} = R_{ik} - \frac{1}{4} g_{ik} R$ is
expanded into bilinear combinations of bivectorial fields being
eigenfunctions of $E$. Field equations for the bivectors induced by Bianchi
identities are studied and it is shown that in general case it will be
3-parametric local symmetry group Yang-Mills field.

\end{abstract}

\sect{Introduction}

It is well known that Einstein equations in General Relativity 
join together pure geometrical quantities in the left side with physical
quantinies (energy-momentum tensor of matter) in the right.

But this fact means that geometry put very rigid restrictions on
energy-momentum tensor and therefore on configurations of all physical fields.
Any permitted mode of physical field has correspondent eigen-mode of
gravitational field otherwise this mode should be prohibited.

We may study geometry types using curvature classifications.
There are two types of curvature classifications: classification of Ricci
tensor by J. Plebansky \cite{pleb} and Petrov classification of Weyl tensor
\cite{pclass}. Both based on studying of eigenvectors of some tensors in
given point of space-time. But eigenvectors of Ricci tensor have not an
immediate physical sense and Weyl tensor types say a little about sources of
gravitational field because it is not affected on Einstein equations.

On the other hand Rainich-Misner-Wheeler already unified theory of
electromagnetic field \cite{es,auth} is not a classification at all. However
it allow to represent curvature of very restricted class of space-times as a
construction of field quantities in finite region of space-time. Meaning of
Rainich conditions is discussed in the second section.

Next section is dedicated to eigenbivectors of irreducible part $E_{iklm}$
of Riemann tensor and its differenial properties. Such approach allows to
generalize already unified theory for sourceless SU(2) Yang-Mills field in
the fourth section.

In the last section the general case of gravitational field sources is
discussed. It is shown that it should be 3-parameric local symmetry group
(maybe noncompact or degenerated) Yang-Mills field with or without sources.

There are five appendicies: on bivectors, on curvature properties, on
electromagnetic energy-momentum tensor structure, on exictance of conformal
transformation provided vanishing of the scalar curvature, and details of
expansive calculations.

In all tensor expressions
latin indices run over (0,1,2,3), greek indices - (1,2,3). Semicolon means
covariant deriviation. 

\sect{Rainich conditions}

If curvature satisfies following conditions

\begin{eqnarray} \label{R0}
R^i_m R^m_k = \frac{1}{4} \delta^i_k R_{mn}R^{mn}, \\ \label{R01} R = 0,
\end{eqnarray}

known as Rainich conditions then it is possible to express irreducible part
of Riemann tensor $E_{iklm}$ defined by equation (\ref{Eiklm}) in following
form

\begin{equation} \label{R1}
E_{iklm} = \frac{1}{2}(f_{ik}f_{lm} + \tilde{f}_{ik}\tilde{f}_{lm}),
\end{equation}

where $f_{ik}$ is bivector and $\tilde{f}_{ik}$ is its dual 
(see Appendix \ref{A}) which satisfy sourceless Maxwell equations
$f^{ik}_{;k}=0, \tilde{f}^{ik}_{;k}=0$.

Contraction of (\ref{R1}) gives

\begin{equation}
S_{ik} = \frac{1}{2}(f_{in}f_k^n+\tilde{f}_{in}\tilde{f}_k^n)
\end{equation}

which is identical with Einstein equation. Really counting (\ref{R01}) there
is Einstein tensor in the left side and
energy-momentum tensor of electromagnetic field in the right.
So we have self-consistent system of electromagnetic and gravitational
field.

It is easy to show (see Appendix \ref{S}) that Rainich conditions (\ref{R0})
and conditions of equality rank of matrix $\mathfrak{S}$ to 1 are the same.

In next section general case of matrix $\mathfrak{S}$ will be studied.

\sect{Eigenbivectors of $E_{iklm}$}

Matrices $A$ and $S$ from (\ref{RAB}) are constructed from vierbein components
of Ricci tensor traceless part $S_{ab}$

\begin{eqnarray}
S &=&
\left( \begin{array}{ccc} 
     S_{11} - S_{00} & S_{12}          & S_{13}          \\
     S_{12}          & S_{22} - S_{00} & S_{23}          \\
     S_{12}          & S_{23}          & S_{33} - S_{00} \\
       \end{array} 
\right),\\
A &=&
\left( \begin{array}{ccc} 
     0       &  S_{03} & -S_{02} \\
    -S_{03}  &  0      &  S_{01} \\
     S_{02}  & -S_{01} &  0      \\
       \end{array} 
\right).
\end{eqnarray}

Let us define $\mathfrak{S} = S - i A$ - hermitian matrix.

Eigenvectors $\mathfrak{F}$ of matrix $\mathfrak{S}$ satisfy equations

\begin{eqnarray*}
  \mathfrak{S} \mathfrak{F} &=& \lambda \mathfrak{F}\\
  E_{iklm} f_{P}^{lm} &=& \lambda f_{ik}
\end{eqnarray*}

Hermitian matrix always have real eigenvalues and it is possible to express
matrix $\mathfrak{S}$ through it eigenvectors

\begin{equation} \label{SM}
   \mathfrak{S}_{\alpha\beta} = \sum_{\iota=1}^{3} \epsilon_{\iota}
  \bidx{\bar{\mathfrak{F}}}{\iota}_{\alpha}\bidx{\mathfrak{F}}{\iota}_{\beta},
\end{equation}

\begin{equation} \label{SE}
   E_{iklm} = \sum_{\iota=1}^{3}
\frac{\epsilon_{\iota}}{2}(\bidx{f}{\iota}_{ik}\bidx{f}{\iota}_{lm} +
                 \bidx{\tilde{f}}{\iota}_{ik}\bidx{\tilde{f}}{\iota}_{lm}),
\end{equation}
\begin{equation} \label{SS}
   S_{ik} = \sum_{\iota=1}^{3}
\frac{\epsilon_{\iota}}{2}(\bidx{f}{\iota}_{ia}\bidx{f}{\iota}^{a}_{k} +
                 \bidx{\tilde{f}}{\iota}_{ia}\bidx{\tilde{f}}{\iota}^{a}_{k}),
\end{equation}

where $\epsilon_{\iota} = sign(\lambda_{\iota}) =
     \left \{ \begin{array}{cc}
              -1 &, \lambda_{\iota} < 0 \\
               0 &, \lambda_{\iota} = 0 \\
               1 &, \lambda_{\iota} > 0
              \end{array}
     \right . $.

$S_{ik}$ looks like energy-momentum tensor of Yang-Mills field with
3 parametric local symmetry group, if the group is compact and
nondegenerated then it is SU(2) or O(3) group.

If scalar curvature $R$ is zero, or if $R$ is nonzero but we applied
conformal transformation described in Appendix \ref{R} then Bianchi
identities (\ref{B2},\ref{B3}) give

\begin{eqnarray}\label{EE}
S^{ik}_{;k} &=& 0\\
C_{iklm}^{;m} &=& E_{iklm}^{;m} = \frac{1}{2} (S_{kl;n} - S_{kn;l}).
\end{eqnarray}

Second equation is consequence of first one, so it is enough to use first
equation.

After substitution $S_{ik}$ from (\ref{SS})

\begin{equation}\label{ED}
\sum_{\iota=1}^3 \epsilon_{\iota} (\bidx{f}{\iota}_{ia}\bidx{f}{\iota}^{ak}_{;k} + 
\bidx{\tilde{f}}{\iota}_{ia}\bidx{\tilde{f}}{\iota}^{ak}) = 0;
\end{equation}

More general expression for divergence $\bidx{f}{\iota}^{ik}_{;k}$ satisfied
equation (\ref{ED}) is

\begin{eqnarray} \label{Div}
\bidx{f}{1}^{ik}_{;k} &=& 
                   - \epsilon_1 \bidx{\tilde{f}}{1}^{ik}\bidx{\xi}{1}_k
                   - \epsilon_2 \bidx{\tilde{f}}{2}^{ik}\bidx{B}{3}_k
                   - \epsilon_3 \bidx{\tilde{f}}{3}^{ik}\bidx{B}{2}_k
                   + \epsilon_2 \bidx{f}{2}^{ik}\bidx{A}{3}_k
                   - \epsilon_3 \bidx{f}{3}^{ik}\bidx{A}{2}_k \\
\bidx{f}{2}^{ik}_{;k} &=& 
                   - \epsilon_2 \bidx{\tilde{f}}{2}^{ik}\bidx{\xi}{2}_k
                   - \epsilon_3 \bidx{\tilde{f}}{3}^{ik}\bidx{B}{1}_k
                   - \epsilon_1 \bidx{\tilde{f}}{1}^{ik}\bidx{B}{3}_k
                   + \epsilon_3 \bidx{f}{3}^{ik}\bidx{A}{1}_k
                   - \epsilon_1 \bidx{f}{1}^{ik}\bidx{A}{3}_k \\
\bidx{f}{3}^{ik}_{;k} &=& 
                   - \epsilon_3 \bidx{\tilde{f}}{3}^{ik}\bidx{\xi}{3}_k
                   - \epsilon_1 \bidx{\tilde{f}}{1}^{ik}\bidx{B}{2}_k
                   - \epsilon_2 \bidx{\tilde{f}}{2}^{ik}\bidx{B}{1}_k
                   + \epsilon_1 \bidx{f}{1}^{ik}\bidx{A}{2}_k
                   - \epsilon_2 \bidx{f}{2}^{ik}\bidx{A}{1}_k \\
\bidx{\tilde{f}}{1}^{ik}_{;k} &=& 
                   + \epsilon_1 \bidx{f}{1}^{ik}\bidx{\xi}{1}_k
                   + \epsilon_2 \bidx{f}{2}^{ik}\bidx{B}{3}_k
                   + \epsilon_3 \bidx{f}{3}^{ik}\bidx{B}{2}_k
                   + \epsilon_2 \bidx{\tilde{f}}{2}^{ik}\bidx{A}{3}_k
                   - \epsilon_3 \bidx{\tilde{f}}{3}^{ik}\bidx{A}{2}_k \\
\bidx{\tilde{f}}{2}^{ik}_{;k} &=& 
                   + \epsilon_2 \bidx{f}{2}^{ik}\bidx{\xi}{2}_k
                   + \epsilon_3 \bidx{f}{3}^{ik}\bidx{B}{1}_k
                   + \epsilon_1 \bidx{f}{1}^{ik}\bidx{B}{3}_k
                   + \epsilon_3 \bidx{\tilde{f}}{3}^{ik}\bidx{A}{1}_k
                   - \epsilon_1 \bidx{\tilde{f}}{1}^{ik}\bidx{A}{3}_k \\
\label{Div2}
\bidx{\tilde{f}}{3}^{ik}_{;k} &=& 
                   + \epsilon_3 \bidx{f}{3}^{ik}\bidx{\xi}{3}_k
                   + \epsilon_1 \bidx{f}{1}^{ik}\bidx{B}{2}_k
                   + \epsilon_2 \bidx{f}{2}^{ik}\bidx{B}{1}_k
                   + \epsilon_1 \bidx{\tilde{f}}{1}^{ik}\bidx{A}{2}_k
                   - \epsilon_2 \bidx{\tilde{f}}{2}^{ik}\bidx{A}{1}_k
\end{eqnarray}

Quantities $A_k$ looks like Yang-Mills potentials, but dependence of
$f_{ik}$ upon $A_k$ is unknown, so they are simply vectorial coefficients.

\sect{Already Unified Theory of SU(2) Yang-Mills field}

Let $\epsilon_{\iota} = 1$, $\xi_k = 0$, $B_k = 0$ then second divergence of
bivectors $f^{ik}$ gives

\begin{eqnarray*}
\bidx{f}{2}^{ik} (\bidx{A}{3}_{k;i} + \bidx{A}{1}_i\bidx{A}{2}_k) &=&
       \bidx{f}{3}^{ik} (\bidx{A}{2}_{k;i} - \bidx{A}{1}_i\bidx{A}{3}_k) \\
\bidx{f}{3}^{ik} (\bidx{A}{1}_{k;i} + \bidx{A}{2}_i\bidx{A}{3}_k) &=&
       \bidx{f}{3}^{ik} (\bidx{A}{2}_{k;i} - \bidx{A}{2}_i\bidx{A}{1}_k) \\
\bidx{f}{1}^{ik} (\bidx{A}{2}_{k;i} + \bidx{A}{3}_i\bidx{A}{1}_k) &=&
       \bidx{f}{2}^{ik} (\bidx{A}{1}_{k;i} - \bidx{A}{3}_i\bidx{A}{2}_k) \\
\end{eqnarray*}

Interpreting these expressions as identities and using antisymmetry of
$f_{ik}$ we obtain usual definitions of SU(2) Yang-Mills field tensors:

\[
f_{ik} = A_{k;i} - A_{i;k} + [A_i,A_k].
\]

Then system of equations (\ref{Div}) becomes

\[
f^{ik}_{;k} + [A_k,f^{ik}] = 0
\]

- sourceless SU(2) Yang-Mills field equations \cite{gauge}.

Einstein equations are already satisfied.

\sect{Field equations in general case}

Now we returning to general case of eigenbivectors. All expansive
calculations are moved into Appendix \ref{calc}.

Second divergence of (\ref{Div}-\ref{Div2}) gives (\ref{FH}-\ref{FH2}).
It is not so easy to express eigenbivectors $f_{ik}$ through their
potentials like in previous section.

Expressions (\ref{FH}-\ref{FH2}) as well as bivectors $\Xi$
(\ref{X}-\ref{X2}) are invariants of gauge group of dual rotation
(\ref{G}-\ref{G2}).

It is possible to fix gauge requiring (\ref{GF}). Such way of gauge fixing
defining 3 new scalar fields $\phi_{\iota}$

\[
\phi_1+\phi_2+\phi_3 = 0.
\]

In this gauge (\ref{FH}-\ref{FH2}) take a form (\ref{FX}-\ref{FX2}).
Now interpreting these equations as identities we obtain expressions for
eigenbivectors. They are consistent only when (\ref{CI}-\ref{CI2}) are true.

Let define
\[
F_{ik} = A_{k;i} - A_{i;k} + [A_i,A_k].
\]

Then first 3 equations of system (\ref{Div}-\ref{Div2}) take a form
\begin{equation} \label{YM1}
F^{ik}_{;k} + [A_k,F^{ik}] = J^i
\end{equation}

of 3-parametric group Yang-Mills field equations.

The last 3 equations of system (\ref{Div}-\ref{Div2}) take a form
\begin{equation} \label{YM2}
\tilde{F}^{ik}_{;k} + [A_k,\tilde{F}^{ik}] = K^i = 0
\end{equation}

these equations with consistency conditions (\ref{CI}-\ref{CI2}) we
interpret as field equations for sources of Yang-Mills field.

Here vectors $J^k$ and $K^k$ are sums of all terms (\ref{Div}-\ref{Div2})
not included into (\ref{YM1},\ref{YM2}) with opposite sign.

\sect{Conclusions}

It is shown that GR Einstein equations allow as a source of the
gravitational field nothing but Yang-Mills field with 3-parametric symmetry
group with or without sources. This means that any other sets of fields must
mimic to demonstrate same behaviour and energy-momentum tensor as
eigen-modes of gravitational field otherwise them will be prohibited.

Nature and properties of sources of Yang-Mills field require more detailed
and careful researches.

\newpage

\begin{appendix}
{\LARGE APPENDICES}

\sect{Bivectors and its vierbein components}
\label{A}

Orthgonal vierbein $h^a_i$ is defined by following expressions:

\begin{eqnarray}
   h_{ia} h^{a}_{k} = g_{ik}; & h^{i}_{a} h_{ib} = \eta_{ab} =
diag \left( \begin{array}{cccc} 1, & -1, & -1, & -1, \end{array} \right),
\end{eqnarray}

where $g_{ik}$ is metrical tensor.

Bivector is an antisymmetric tensor $f_{ik} = - f_{ki}$. Vierbien components of
bivector $f_{ab} = h^i_a h^k_b f_{ik}$ 

\[
   f_{ab} = 
\left( \begin{array}{cccc} 
     0       &  e_{1} &  e_{2} &  e_{3} \\
    -e_{1}   &  0     & -h_{3} &  h_{2} \\
    -e_{2}   &  h_{3} &  0     & -h_{1} \\
    -e_{3}   & -h_{2} &  h_{1} &  0     \\
       \end{array} 
\right).
\]

Using usual remapping of bivector indices

\begin{center}
\begin{tabular}{|c|c|c|c|c|c|c|}\hline
 A  & 1 &  2 &  3 &  4 &  5 &  6\\\hline
ik & 01 & 02 & 03 & 32 & 13 & 21\\\hline
\end{tabular}
\end{center}
\vskip 2mm

it is possible to write same bivecor as real 6-vector or as complex 3-vector

$F = (e_{1},e_{2},e_{3},h_{1},h_{2},h_{3})$,
$\mathfrak{F} = ( e_{1} + i h_{1}, e_{2} + i h_{2}, e_{3} + i h_{3})$.

Dual bivector defined as

\[
\tilde{f}_{ik} \equiv \frac{\sqrt{-g}}{2} \epsilon_{iklm} f^{lm},
\]

where $g$ is determinant of metrical tensor $g_{ik}$ and $\epsilon_{iklm}$
is absolutely antisymmetric Levi-Civita pseudotensor, has components

\[
  \tilde{f}_{ab} = 
\left( \begin{array}{cccc} 
     0       & -h_{1} & -h_{2} & -h_{3} \\
     h_{1}   &  0     & -e_{3} &  e_{2} \\
     h_{2}   &  e_{3} &  0     & -e_{1} \\
     h_{3}   & -e_{2} &  e_{1} &  0     \\
       \end{array} 
\right)
\]

$\tilde{F} = (-h_{1},-h_{2},-h_{3},e_{1},e_{2},e_{3})$,
$\tilde{\mathfrak{F}} = 
( -h_{1} + i e_{1}, -h_{2} + i e_{2}, -h_{3} + i e_{3})$.

\vskip 2mm

Useful identity for bivectors $X_{ik}$ and $Y_{lm}$

\[
  X_{ia} Y^{\;a}_{k} - \tilde{X}_{ka} \tilde{Y}^{\;a}_{i} = \frac{1}{2} g_{ik} X_{ab} Y^{ab}.
\]

It is possible to define so-called dual rotations with parameter $\varphi$

\[
    f_{ik} \rightarrow 
                f_{ik}  \cos{\varphi} - \tilde{f}_{ik} \sin{\varphi},
\]
\[
    \tilde{f}_{ik} \rightarrow 
                  f_{ik}  \sin{\varphi} + \tilde{f}_{ik} \cos{\varphi}.
\]

Vierbein components of parity conjugated contravariant bivector are the same
as covariant vierbein components of original one:

\[
   P f^{ab} = f_{P}^{ab} =
\left( \begin{array}{cccc} 
     0       &  e_{1} &  e_{2} &  e_{3} \\
    -e_{1}   &  0     & -h_{3} &  h_{2} \\
    -e_{2}   &  h_{3} &  0     & -h_{1} \\
    -e_{3}   & -h_{2} &  h_{1} &  0     \\
       \end{array} 
\right).
\]

Contraction of any selfdual bivector 
$f^{(+)}_{ik} \equiv f_{ik} - i \tilde{f}_{ik}$
with any antiselfdual bivector
$g^{(-)}_{ik} \equiv g_{ik} + i \tilde{g}_{ik}$ is zero
$f^{(+)}_{ik}g^{(-)ik} = 0$.

\sect{Curvature tensor and its properties}

Riemann tensor defined as
\[
R^i_{klm} = \frac{\partial \Gamma^i_{km}}{\partial x^l} - 
            \frac{\partial \Gamma^i_{kl}}{\partial x^m} + 
       \Gamma^i_{nl}\Gamma^n_{km} - \Gamma^i_{nm}\Gamma^n_{kl},
\]

where $\Gamma^i_{nl} = \frac{1}{2}g^{ij}(\frac{\partial g_{kj}}{x^l} +
\frac{\partial g_{jl}}{x^k} - \frac{\partial g_{kl}}{x^j})$ 
is a Christoffel symbol of the second kind.

\subsection{Algebraic properties}

Riemann tensor has following symmetries:

\begin{eqnarray*}
R_{iklm} &=& - R_{kilm} = - R_{ikml} \\
R_{iklm} &=& R_{lmik} \\
R_{iklm} &+& R_{imkl} + R_{ilmk} = 0,
\end{eqnarray*}

so it has 20 indepenent components.

Contractions of Riemann tensor are known as Ricci tensor and scalar
curvature:

\begin{eqnarray*}
R_{ik} = R^l_{ilk}, & R_{ik} = R_{ki} \\
\end{eqnarray*}
\[
R = R^i_i
\]

Using bivectorial remapping of first and second indices pairs of Riemann
tensor it is possible to rewrite it as symmetric 6x6 matrix

\begin{equation} \label{RAB}
R_{iklm} \to R_{AB} = R_{BA} =
\left( \begin{array}{cc} M &  N \\  N & -M \end{array} \right) +
\left( \begin{array}{cc} S &  A \\ -A &  S \end{array} \right),
\end{equation}
where $M, N, S, A$ - 3x3 matrices and
\[
   M_{\alpha\beta} = M_{\beta\alpha},\; N_{\alpha\beta} = N_{\beta\alpha},\;
   S_{\alpha\beta} = S_{\beta\alpha},\; A_{\alpha\beta} = -A_{\beta\alpha},
\]
$A, B$ = 1..6;  $\alpha, \beta$ = 1..3.
\[
M_{11}+M_{22}+M_{33} = \frac{R}{2};\ \ \  N_{11} + N_{22} + N_{33} = 0;
\]

Riemann tensor is expandible into following irreducible parts

\begin{equation} \label{IRR}
R_{iklm} = C_{iklm} + E_{iklm} + G_{iklm},
\end{equation}
where $C_{iklm}$ is so-called conformaly invariant Weyl tensor and
\begin{equation} \label{Eiklm}
E_{iklm} = \frac{1}{2} ( g_{il}S_{km} + g_{km}S_{il} - g_{im}S_{kl} -
                                                           g_{kl}S_{im} );
\end{equation}
\begin{equation} \label{Giklm}
G_{iklm} = \frac{R}{12} ( g_{il}g_{km} - g_{im}g_{kl} );
\end{equation}

$S_{ik} \equiv R_{ik} - \frac{R}{4}g_{ik}$ - Ricci tensor traceless part.

Matrices $M$ and $N$ of \ref{RAB} are constructed from components of Weyl
tensor $C_{iklm}$ and scalar curvature $R$ and matrices $A$ and $S$ - from
components of $E_{iklm}$ (or $S_{ik}$).

\subsection{Differential properties}

Riemann tensor satisfies Bianchi identities

\begin{equation} \label{B1}
R^n_{ikl;m} + R^n_{imk;l} + R^n_{ilm;k} = 0,
\end{equation}

and contracted Bianchi identities

\begin{equation} \label{B2}
R^m_{ikl;m} + R_{ik;l} - R_{il;k} = 0,
\end{equation}
\begin{equation} \label{B3}
(R^i_k - \frac{1}{2}R \delta^i_k)_{;i} = 0.
\end{equation}

\sect{Structure of electromagnetic field energy-momentum tensor}
\label{S}

Energy-momentum tensor of electromagnetic field is defined by following
expression:
\[
  T_{ik} = - f_{ia} f_{k}^{a} + \frac{1}{4} g_{ik} f_{ab} f^{ab}
         = - \frac{1}{2}( f_{ia} f_{k}^{a} + \widetilde{f_{ia}} \widetilde{f_{k}^{a}} ).
\]

It is possible to express its vierbein components through electromagnetic
field components either in real bivector form or in complex 3-dimensional
vector
$\mathfrak{F} = ( e_{1} + i h_{1}, e_{2} + i h_{2}, e_{3} + i h_{3})$
and complex conjugated vector
$\bar{\mathfrak{F}} = ( e_{1} - i h_{1}, e_{2} - i h_{2}, e_{3} - i h_{3})$
following way:

\begin{eqnarray*}
  T_{00} =& \frac{1}{2}(e_1^2 + e_2^2 + e_3^2 + h_1^2 + h_2^2 + h_3^2)
         &=  \frac{1}{2}(\bar{\mathfrak{F}}_1 \mathfrak{F}_1 + 
                       \bar{\mathfrak{F}}_2 \mathfrak{F}_2  + 
                       \bar{\mathfrak{F}}_3 \mathfrak{F}_3 ),\\
  T_{11} =& \frac{1}{2}(-e_1^2 + e_2^2 + e_3^2 - h_1^2 + h_2^2 + h_3^2)
         &=  \frac{1}{2}(-\bar{\mathfrak{F}}_1 \mathfrak{F}_1 + 
                       \bar{\mathfrak{F}}_2 \mathfrak{F}_2  + 
                       \bar{\mathfrak{F}}_3 \mathfrak{F}_3 ),\\
  T_{22} =& \frac{1}{2}(e_1^2 - e_2^2 + e_3^2 + h_1^2 - h_2^2 + h_3^2)
         &=  \frac{1}{2}(\bar{\mathfrak{F}}_1 \mathfrak{F}_1 -
                       \bar{\mathfrak{F}}_2 \mathfrak{F}_2  + 
                       \bar{\mathfrak{F}}_3 \mathfrak{F}_3 ),\\
  T_{33} =& \frac{1}{2}(e_1^2 + e_2^2 - e_3^2 + h_1^2 + h_2^2 - h_3^2)
         &=  \frac{1}{2}(\bar{\mathfrak{F}}_1 \mathfrak{F}_1 + 
                       \bar{\mathfrak{F}}_2 \mathfrak{F}_2  - 
                       \bar{\mathfrak{F}}_3 \mathfrak{F}_3 ),\\
  T_{01} =&            - e_2 h_3 + h_2 e_3 
         &=  \frac{i}{2}(\bar{\mathfrak{F}}_2 \mathfrak{F}_3 - 
                       \bar{\mathfrak{F}}_3 \mathfrak{F}_2 ),\\
  T_{02} =&              e_1 h_3 - h_1 e_3 
         &=  \frac{i}{2}(-\bar{\mathfrak{F}}_1 \mathfrak{F}_3 + 
                       \bar{\mathfrak{F}}_3 \mathfrak{F}_1 ),\\
  T_{03} =&            - e_1 h_2 + h_1 e_2 
         &=  \frac{i}{2}(\bar{\mathfrak{F}}_1 \mathfrak{F}_2 - 
                       \bar{\mathfrak{F}}_2 \mathfrak{F}_1 ),\\
  T_{12} =&            - e_1 e_2 - h_1 h_2 
         &= - \frac{1}{2}(\bar{\mathfrak{F}}_1 \mathfrak{F}_2 + 
                       \bar{\mathfrak{F}}_2 \mathfrak{F}_1 ),\\
  T_{13} =&            - e_1 e_3 - h_1 h_3 
         &= - \frac{1}{2}(\bar{\mathfrak{F}}_1 \mathfrak{F}_3 + 
                       \bar{\mathfrak{F}}_3 \mathfrak{F}_1 ),\\
  T_{23} =&            - e_2 e_3 - h_2 h_3 
         &= - \frac{1}{2}(\bar{\mathfrak{F}}_2 \mathfrak{F}_3 + 
                       \bar{\mathfrak{F}}_3 \mathfrak{F}_2 ).
\end{eqnarray*}

It is evident that previous formulae are expressible in 3x3 hermitian matrix
form:

\begin{equation*}
\mathfrak{S} =
\left( \begin{array}{ccc} 
     T_{11} -   T_{00} & T_{12} + i T_{03} & T_{13} - i T_{02} \\
     T_{12} - i T_{03} & T_{22} -   T_{00} & T_{23} + i T_{01} \\
     T_{12} + i T_{02} & T_{23} - i T_{01} & T_{33} -   T_{00} \\
       \end{array} 
\right) = -
\left( \begin{array}{ccc} 
    \bar{\mathfrak{F}}_1 \mathfrak{F}_1  & \bar{\mathfrak{F}}_1 \mathfrak{F}_2 & \bar{\mathfrak{F}}_1 \mathfrak{F}_3 \\
    \bar{\mathfrak{F}}_2 \mathfrak{F}_1  & \bar{\mathfrak{F}}_2 \mathfrak{F}_2 & \bar{\mathfrak{F}}_2 \mathfrak{F}_3 \\
    \bar{\mathfrak{F}}_3 \mathfrak{F}_1  & \bar{\mathfrak{F}}_3 \mathfrak{F}_2 & \bar{\mathfrak{F}}_3 \mathfrak{F}_3 
       \end{array} 
\right).
\end{equation*}

Matrix $\mathfrak{S}$ has a rank 1 i.e. all its subdeterminants are zero.
It is easy to prove that former statement is equivalent to so-called Rainich
conditions \cite{es,auth}:

\[
   T_{ia} T_{k}^{a} = \frac{1}{4} g_{ik} T_{ab} T^{ab}.
\]

\sect{ On existence of conformal transformation provided vanishing of the
scalar curvature}
\label{R}

Let given Riemannian space $V_4$ with metric $g_{ik}$, Riemann tensor
$R_{iklm}$, Ricci tensor $R_{ik} = R^{a}_{iak}$ and scalar 
curvarure $R = R^{a}_{a} \not\equiv 0$. We shell find conformal
transformation
\begin{eqnarray*}
g_{ik} &\rightarrow& \bar{g}_{ik} = \varphi g_{ik}, \\
R_{iklm} &\rightarrow& \bar{R}_{iklm}, \\
R_{ik} &\rightarrow& \bar{R}_{ik}, \\
R &\rightarrow& \bar{R} = 0, \\
\end{eqnarray*}

which provides vanishing of $\bar{R}$. Riemann tensor of conformal metric is

\begin{eqnarray*}
  \bar{R}_{iklm} = \varphi R_{iklm}
       &+& \frac{1}{2}     (g_{im}\varphi_{kl} + g_{kl}\varphi_{im} - 
                          g_{il}\varphi_{km} - g_{km}\varphi_{il}) \\
       &-& \frac{3}{4\varphi} (g_{im}\varphi_{k}\varphi_{l} + g_{kl}\varphi_{i}\varphi_{m} -
                          g_{il}\varphi_{k}\varphi_{m} - g_{km}\varphi_{i}\varphi_{l}) \\
       &+& \frac{1}{4\varphi} (g_{im}g_{kl} - g_{km}g_{il}) \varphi_{n}\varphi^{n}, 
\end{eqnarray*}

where $\varphi_{i} \equiv \nabla_{i} \varphi$ 
É $\varphi_{ik} \equiv \nabla_{i}\nabla_{k} \varphi$. Then
\begin{eqnarray*}
  \bar{R}_{ik} &=& R_{ik} - \frac{\varphi_{ik}}{\varphi} 
                          - \frac{1}{2\varphi} g_{ik}\nabla_{n}\nabla^{n}\varphi
                          + \frac{3}{2\varphi^{2}} \varphi_{i}\varphi_{k}, \\
  \bar{R}      &=& R      - \frac{3}{\varphi} \nabla_{n}\nabla^{n}\varphi
                          + \frac{3}{2\varphi^{2}} \varphi_{n}\varphi^{n}.
\end{eqnarray*}

Equating $\bar{R}$ to zero and making substitution $\varphi = \psi^2$
we obtain so-called conformal scalar field equation \cite{conf}:

\[
  \nabla_{i}\nabla^{i} \psi - \frac{1}{6} R \psi = 0.
\]

\sect{Detailed calculations}
\label{calc}

Let us introduce complex field variables to reduce expressions.

\begin{eqnarray}\label{C}
\bidx{\mathfrak{A}}{\iota}_{i} &=& \bidx{A}{\iota}_{i} + i \bidx{B}{\iota}_{i} \\
\bidx{\mathfrak{F}}{\iota}_{ik} &=& \bidx{f}{\iota}_{ik} + i \bidx{\tilde{f}}{\iota}_{ik} \\
\bidx{\mathfrak{H}}{\iota}_{ik} &=& \bidx{\Phi}{\iota}_{ik} + i \bidx{\Theta}{\iota}_{ik}
\end{eqnarray}

so $\tilde{\mathfrak{F}} = -i \mathfrak{F}$.

Then (\ref{Div}-\ref{Div2}) becomes

\begin{eqnarray}\label{F}
\bidx{\mathfrak{F}}{1}^{ik}_{;k} &=& 
    i \epsilon_1 \bidx{\mathfrak{F}}{1}^{ik} \bidx{\xi}{1}_{k} +
      \epsilon_2 \bidx{\mathfrak{F}}{2}^{ik} \bidx{\mathfrak{A}}{3}_{k} -
      \epsilon_3 \bidx{\mathfrak{F}}{3}^{ik} \bidx{\mathfrak{A}}{2}^{*}_{k} \\
\bidx{\mathfrak{F}}{2}^{ik}_{;k} &=& 
    i \epsilon_2 \bidx{\mathfrak{F}}{2}^{ik} \bidx{\xi}{2}_{k} +
      \epsilon_3 \bidx{\mathfrak{F}}{3}^{ik} \bidx{\mathfrak{A}}{1}_{k} -
      \epsilon_1 \bidx{\mathfrak{F}}{1}^{ik} \bidx{\mathfrak{A}}{3}^{*}_{k} \\
\bidx{\mathfrak{F}}{3}^{ik}_{;k} &=& 
    i \epsilon_3 \bidx{\mathfrak{F}}{1}^{ik} \bidx{\xi}{3}_{k} +
      \epsilon_1 \bidx{\mathfrak{F}}{1}^{ik} \bidx{\mathfrak{A}}{2}_{k} -
      \epsilon_2 \bidx{\mathfrak{F}}{2}^{ik} \bidx{\mathfrak{A}}{1}^{*}_{k}
\end{eqnarray}

where ${}^{*}$ means complex conjugation.

Let introduce complex bivectorial field $\mathfrak{H}$

\begin{eqnarray}\label{H}
\bidx{\mathfrak{H}}{1}^{ik} &=& 
      \bidx{\mathfrak{A}}{1}_{[k;i]} +
      \epsilon_1 (\bidx{\mathfrak{A}}{2}^{*}_{[i} \bidx{\mathfrak{A}}{3}^{*}_{k]} -
      i \bidx{\xi}{1}^{'}_{[i} \bidx{\mathfrak{A}}{1}_{k]}) \\
\bidx{\mathfrak{H}}{2}^{ik} &=& 
      \bidx{\mathfrak{A}}{2}_{[k;i]} +
      \epsilon_2 (\bidx{\mathfrak{A}}{3}^{*}_{[i} \bidx{\mathfrak{A}}{1}^{*}_{k]} -
      i \bidx{\xi}{2}^{'}_{[i} \bidx{\mathfrak{A}}{2}_{k]}) \\
\bidx{\mathfrak{H}}{3}^{ik} &=& 
      \bidx{\mathfrak{A}}{3}_{[k;i]} +
      \epsilon_3 (\bidx{\mathfrak{A}}{1}^{*}_{[i} \bidx{\mathfrak{A}}{2}^{*}_{k]} -
      i \bidx{\xi}{3}^{'}_{[i} \bidx{\mathfrak{A}}{3}_{k]})
\end{eqnarray}

where $[$ $]$ means alternation,

\begin{eqnarray}\label{x}
\epsilon_1 \bidx{\xi}{1}^{'} = \epsilon_2 \bidx{\xi}{2} - \epsilon_3 \bidx{\xi}{3}\\
\epsilon_2 \bidx{\xi}{2}^{'} = \epsilon_3 \bidx{\xi}{3} - \epsilon_1 \bidx{\xi}{1}\\
\epsilon_3 \bidx{\xi}{3}^{'} = \epsilon_1 \bidx{\xi}{1} - \epsilon_2 \bidx{\xi}{2}
\end{eqnarray}

And real field $\Xi$

\begin{eqnarray}\label{X}
\bidx{\Xi}{1}_{ik} &=& \bidx{\xi}{1}_{[k;i]} - 
         2 \epsilon_2 \bidx{A}{3}_{[i} \bidx{B}{3}_{k]} + 
         2 \epsilon_3 \bidx{A}{2}_{[i} \bidx{B}{2}_{k]} \\
\bidx{\Xi}{2}_{ik} &=& \bidx{\xi}{2}_{[k;i]} - 
         2 \epsilon_3 \bidx{A}{1}_{[i} \bidx{B}{1}_{k]} + 
         2 \epsilon_1 \bidx{A}{3}_{[i} \bidx{B}{3}_{k]} \\
\label{X2}
\bidx{\Xi}{3}_{ik} &=& \bidx{\xi}{3}_{[k;i]} - 
         2 \epsilon_1 \bidx{A}{2}_{[i} \bidx{B}{2}_{k]} + 
         2 \epsilon_2 \bidx{A}{1}_{[i} \bidx{B}{1}_{k]}
\end{eqnarray}

Due to vanishing of second divergence of any bivector

\begin{eqnarray}\label{FH}
\epsilon_2 \bidx{\mathfrak{F}}{2}^{ik}\bidx{\mathfrak{H}}{3}_{ik} - 
    \epsilon_3 \bidx{\mathfrak{F}}{3}^{ik}\bidx{\mathfrak{H}}{2}^{*}_{ik} + 
  i \epsilon_1 \bidx{\mathfrak{F}}{1}^{ik}\bidx{\Xi}{1}_{ik}=0\\
\epsilon_3 \bidx{\mathfrak{F}}{3}^{ik}\bidx{\mathfrak{H}}{1}_{ik} - 
    \epsilon_1 \bidx{\mathfrak{F}}{1}^{ik}\bidx{\mathfrak{H}}{3}^{*}_{ik} + 
  i \epsilon_2 \bidx{\mathfrak{F}}{2}^{ik}\bidx{\Xi}{2}_{ik}=0\\
\label{FH2}
\epsilon_1 \bidx{\mathfrak{F}}{1}^{ik}\bidx{\mathfrak{H}}{2}_{ik} - 
    \epsilon_2 \bidx{\mathfrak{F}}{2}^{ik}\bidx{\mathfrak{H}}{1}^{*}_{ik} + 
  i \epsilon_3 \bidx{\mathfrak{F}}{3}^{ik}\bidx{\Xi}{3}_{ik}=0
\end{eqnarray}

Transformations of the fields under dual rotations

\begin{eqnarray}\label{G}
\bidx{\mathfrak{F}}{\iota} &\to&
e^{-i\epsilon_{\iota}\alpha_{\iota}}\bidx{\mathfrak{F}}{\iota}\\
\bidx{\mathfrak{A}}{\iota} &\to&
e^{-i\epsilon_{\iota}\alpha^{'}_{\iota}}\bidx{\mathfrak{A}}{\iota}\\
\label{G2}
\bidx{\mathfrak{H}}{\iota} &\to&
e^{-i\epsilon_{\iota}\alpha^{'}_{\iota}}\bidx{\mathfrak{H}}{\iota}
\end{eqnarray}

where

\begin{eqnarray}\label{a}
\epsilon_1\bidx{\alpha}{1}^{'} = \epsilon_2 \bidx{\alpha}{2} - \epsilon_3 \bidx{\alpha}{3}\\
\epsilon_2\bidx{\alpha}{2}^{'} = \epsilon_3 \bidx{\alpha}{3} - \epsilon_1 \bidx{\alpha}{1}\\
\epsilon_3\bidx{\alpha}{3}^{'} = \epsilon_1 \bidx{\alpha}{1} - \epsilon_2 \bidx{\alpha}{2}
\end{eqnarray}

$\Xi$ is invariant under dual rotations. It is evident that equations
(\ref{FH}-\ref{FH2}) are also invariant.

Let

\begin{eqnarray}\label{scalar}
\frac{\varphi_{2}}{\varphi_{3}} = e^{\phi_{1}}, \;\;
\frac{\varphi_{3}}{\varphi_{1}} = e^{\phi_{2}}, \;\;
\frac{\varphi_{1}}{\varphi_{2}} = e^{\phi_{3}},
\end{eqnarray}

\[
\phi_1+\phi_2+\phi_3 = 0;
\]

where $\varphi_{\iota}$ are arbitrary positive real scalar functions.

To solve (\ref{FH}-\ref{FH2}) if is enough to fix gauge requiring

\begin{equation} \label{GF}
  \epsilon_1 \varphi_1\bidx{\mathfrak{F}}{1}^{ik}\bidx{\Xi}{1}_{ik} +
  \epsilon_2 \varphi_2\bidx{\mathfrak{F}}{2}^{ik}\bidx{\Xi}{2}_{ik} +
  \epsilon_3 \varphi_3\bidx{\mathfrak{F}}{3}^{ik}\bidx{\Xi}{3}_{ik}=0
\end{equation}

Then

\begin{eqnarray}\label{FX}
\epsilon_2 \bidx{\mathfrak{F}}{2}^{ik}
        (\bidx{\mathfrak{H}}{3}_{ik} - i e^{-\phi_3} \bidx{\Xi}{2}_{ik}) &=&
    \epsilon_3 \bidx{\mathfrak{F}}{3}^{ik}
        (\bidx{\mathfrak{H}}{2}^{*}_{ik} + i e^{\phi_2} \bidx{\Xi}{3}_{ik})\\
\epsilon_3 \bidx{\mathfrak{F}}{3}^{ik}
        (\bidx{\mathfrak{H}}{1}_{ik} - i e^{-\phi_1} \bidx{\Xi}{3}_{ik}) &=&
    \epsilon_1 \bidx{\mathfrak{F}}{1}^{ik}
        (\bidx{\mathfrak{H}}{3}^{*}_{ik} + i e^{\phi_3} \bidx{\Xi}{1}_{ik})\\
\label{FX2}
\epsilon_1 \bidx{\mathfrak{F}}{1}^{ik}
        (\bidx{\mathfrak{H}}{2}_{ik} - i e^{-\phi_2} \bidx{\Xi}{1}_{ik}) &=&
    \epsilon_2 \bidx{\mathfrak{F}}{2}^{ik}
        (\bidx{\mathfrak{H}}{1}^{*}_{ik} + i e^{\phi_1} \bidx{\Xi}{2}_{ik})
\end{eqnarray}

So

\begin{eqnarray}\label{FY}
\epsilon_1 \bidx{\mathfrak{F}}{1}_{ik} &=
 \bidx{\mathfrak{H}}{1}_{ik} - i e^{-\phi_1} \bidx{\Xi}{3}_{ik} &=
 \bidx{\mathfrak{H}}{1}^{*}_{ik} + i e^{\phi_1} \bidx{\Xi}{2}_{ik}\\
\epsilon_2 \bidx{\mathfrak{F}}{2}_{ik} &=
 \bidx{\mathfrak{H}}{2}_{ik} - i e^{-\phi_2} \bidx{\Xi}{1}_{ik} &=
 \bidx{\mathfrak{H}}{2}^{*}_{ik} + i e^{\phi_2} \bidx{\Xi}{3}_{ik}\\
\label{FY2}
\epsilon_3 \bidx{\mathfrak{F}}{3}_{ik} &=
 \bidx{\mathfrak{H}}{3}_{ik} - i e^{-\phi_3} \bidx{\Xi}{2}_{ik} &=
 \bidx{\mathfrak{H}}{3}^{*}_{ik} + i e^{\phi_3} \bidx{\Xi}{1}_{ik}
\end{eqnarray}

Set of consistency conditions of system (\ref{FY}-\ref{FY2}) is

\begin{eqnarray}\label{CI}
\bidx{\Theta}{1}_{ik} &=
 \frac{e^{\phi_1}}{2} \bidx{\Xi}{2}_{ik} + \frac{e^{-\phi_1}}{2} \bidx{\Xi}{3}_{ik}\\
\bidx{\Theta}{2}_{ik} &=
 \frac{e^{\phi_2}}{2} \bidx{\Xi}{3}_{ik} + \frac{e^{-\phi_2}}{2} \bidx{\Xi}{1}_{ik}\\
\bidx{\Theta}{3}_{ik} &=
 \frac{e^{\phi_3}}{2} \bidx{\Xi}{1}_{ik} + \frac{e^{-\phi_3}}{2} \bidx{\Xi}{2}_{ik}
\\
%
%
\bidx{\tilde{\Phi}}{1}_{ik} &=
 \frac{e^{\phi_1}}{2} \bidx{\Xi}{2}_{ik} - \frac{e^{-\phi_1}}{2} \bidx{\Xi}{3}_{ik}\\
\bidx{\tilde{\Phi}}{2}_{ik} &=
 \frac{e^{\phi_2}}{2} \bidx{\Xi}{3}_{ik} - \frac{e^{-\phi_2}}{2} \bidx{\Xi}{1}_{ik}\\
\label{CI2}
\bidx{\tilde{\Phi}}{3}_{ik} &=
 \frac{e^{\phi_3}}{2} \bidx{\Xi}{1}_{ik} - \frac{e^{-\phi_3}}{2} \bidx{\Xi}{2}_{ik}
\end{eqnarray}

Now

\begin{eqnarray}
\epsilon_1 \bidx{f}{1}_{ik} &= \bidx{\Phi}{1}_{ik} &= 
\bidx{A}{1}_{[k;i]} + \epsilon_1 (\bidx{A}{2}_{[i}\bidx{A}{3}_{k]} -
\bidx{B}{2}_{[i}\bidx{B}{3}_{k]} + \bidx{\xi}{1}^{'}_{[i} \bidx{B}{1}_{k]}) \\
\epsilon_2 \bidx{f}{2}_{ik} &= \bidx{\Phi}{2}_{ik} &= 
\bidx{A}{2}_{[k;i]} + \epsilon_2 (\bidx{A}{3}_{[i}\bidx{A}{1}_{k]} -
\bidx{B}{3}_{[i}\bidx{B}{1}_{k]} + \bidx{\xi}{2}^{'}_{[i} \bidx{B}{2}_{k]}) \\
\epsilon_3 \bidx{f}{3}_{ik} &= \bidx{\Phi}{3}_{ik} &= 
\bidx{A}{3}_{[k;i]} + \epsilon_3 (\bidx{A}{1}_{[i}\bidx{A}{2}_{k]} -
\bidx{B}{1}_{[i}\bidx{B}{2}_{k]} + \bidx{\xi}{3}^{'}_{[i} \bidx{B}{3}_{k]})
\end{eqnarray}

\end{appendix}

\end{document}